# Physics-Based Models for Magneto-Electric Spin-Orbit Logic Circuits

Hai Li, Dmitri E. Nikonov, *Senior Member, IEEE*, Chia-Ching Lin, Kerem Camsari, Yu-Ching Liao, *Student Member, IEEE*, Chia-Sheng Hsu, *Student Member, IEEE*, Azad Naeemi, *Senior Member, IEEE*, and Ian A. Young, *Fellow, IEEE*

1

***Abstract*—Spintronic devices are a promising beyond-CMOS device option thanks to their energy efficiency and compatibility with CMOS. To accurately capture their multi-physics dynamics, a rigorous treatment of both spin and charge and their inter-conversion is required. Here we present physics-based device models based on 4x4 matrices for the spin-orbit coupling part of the magneto-electric spin-orbit (MESO) device. Also, a more rigorous physics model of ferroelectric and magnetoelectric switching of ferromagnets, based on Landau-Lifshitz-Gilbert (LLG) and Landau-Khalatnikov (LK) equations, is presented. With the combined model implemented in a SPICE circuit simulator environment, simulation results were obtained which show feasibility of MESO implementation and functional operation of buffers, oscillators, and majority gates.**

***Index Terms*— Beyond CMOS logic, magnetoelectric, spin-orbit, spintronic devices, SPICE.**

## I. Simulations of Beyond CMOS Circuits

THE scaling of integrated circuits based on CMOS transistors has been carried out over the last four decades in agreement with the Moore's law [1], [2]. In the last 15 years, power density is becoming a more critical limitation [3]. To address this issue and to assure continued scaling, beyond CMOS devices have been explored. They are intended to complement CMOS and be monolithically integrated in the same die. Compared with traditional CMOS-based circuits, new computing variables – ferroelectric polarization and magnetization – were utilized in beyond CMOS devices. Via systematic benchmarking [4], spintronic devices were found to be a promising option with better energy efficiency. To describe such spintronics device based circuits, a method has been developed, where 4x4 matrices represent various spintronic components [5]–[9]. 4x4 matrices relate 4-component current (3 spin + 1 charge) to 4-component voltage. To utilize the potential of spintronics for saving energy in computing, various logic devices have been proposed. Among them, devices involving magnetoelectric effects, such as MESO [10], [11] CoMET [12], ASFOR [13], and SOTFET [14], are expected to operate with lower energy.

In this work, we build on the previous models of MESO [10], [11], [15]. Here we replace the approximated compact models with models based on more rigorous physics of their switching. Then we use these device models to construct a few of the representative circuits. This model development enables a more rigorous and precise analysis of the tight coupling and interplay of multiple nontrivial physics phenomena involved in MESO operation. Specifically, 4x4 matrices derived from the drift-diffusion equations are used to solve the spin-charge conversion in the SO output of the MESO device . In the ME input part of MESO, the dynamics of coupled ferroelectric-antiferromagnetic-ferromagnetic (FE-AFM-FM) layers is captured with both the Landau-Lifshitz-Gilbert (LLG) equation and the Landau-Khalatnikov (LK) equation [16]. Meanwhile, since all aspects of the device and circuit are simulated in a SPICE environment [17], convenient design and technology co-optimization (DTCO) are enabled across materials, devices, and circuits.

## II. Limitations of Prior Treatment of Spintronic Circuits

In a recent MESO model update, we proposed to modify the original single-sided MESO device into a differential one [15]. Fig. 1 shows cascading of two differential MESO devices, where a vertical blue plane designates a cross-section through the device's SO output part. Compared with the single-sided version, there are two primary changes. Firstly, both outputs ($+V_{out}$ /$-V_{out}$) of the spin-orbit (SO) part in the 1st MESO form a differential signal and are connected to the top and bottom electrodes of magnetoelectric (ME) capacitor in 2nd MESO, respectively. Secondly, an electrically insulating layer (gray), but magnetically coupled, is inserted between two ferromagnetic layers (red). These changes brought multiple

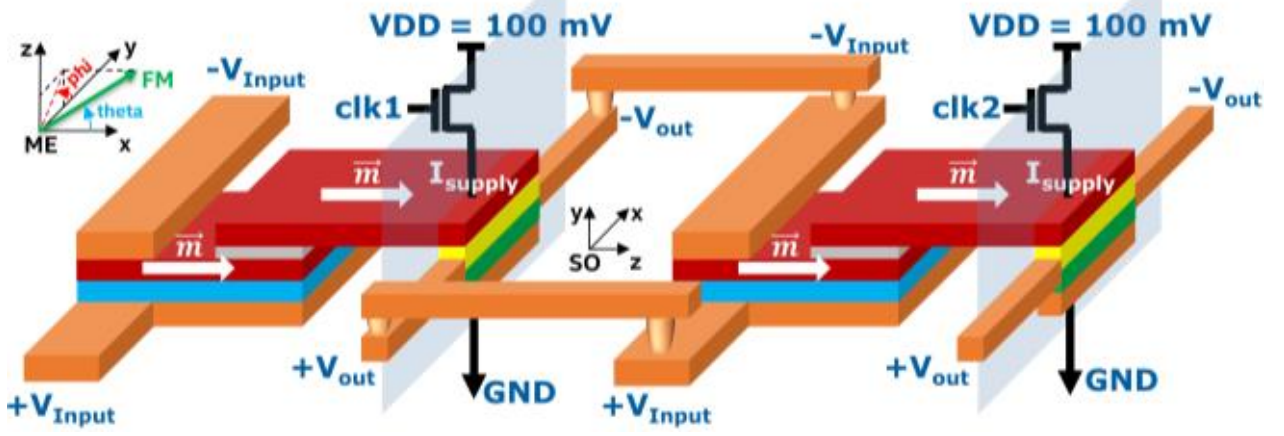

*Figure 1. A side view of two cascaded stages of differential MESO. The 'slice' through the SO part is shown by a blue vertical plane. $+/-V_{in}$ and $+/-V_{out}$ are two pairs of input/output terminals. Coordinates in 2 insets.*

H. Li, D. E. Nikonov, C.-C. Lin, and I. A. Young are with Components Research, Intel Corp., Hillsboro, OR 97124, USA (e-mail: hai.li@intel.com).

K. Camsari is with University of California, Santa Barbara, Y.-C. Liao, C.-S. Hsu, and A. Naeemi are with the Georgia Institute of Technology.

advantages: (1) Enablement of differential input/output signals; (2) prevention of interference between adjacent MESO devices; (3) elimination of the footer transistor in each stage; (4) simplification of clocking control from multiple overlapping clocks to just two non-overlapping clocks.

The equivalent circuit model for two cascaded differential MESO is shown in Fig. 2. With differential inputs/outputs and with an insulating layer, each blue-dash-box (MESO #1 for instance) represents an isolated circuit piece. The branch including nodes $s1_1$-$c_1$-$s2_1$ (subscript number indicates MESO device number #1 or #2) is the ME input part, where resistance $R_{FM}$ and capacitance $C_{FE}$ are in series. These two elements are used to represent the delay due to ferromagnetic (FM) layer (top electrode of $C_{ME}$) and the ferroelectric capacitor of multiferroic layer (blue layer in Fig. 1). The other branch consists of the vertical path of $V_{DD}$-$t_1$-gnd and the loop of $t_1$-$a_1$-$o_1$-$u_1$-$b_1$-$t_1$. $R_{S1}$ and $R_{S2}$ approximate the resistance of the layer stack that vertical current traverses. In the loop, $R_{ISOC}$ is split into two halves, each with $I_{SOC}$ in parallel. This is to capture the charge-spin-charge conversion from vertical path to the loop. The interconnect resistance is simulated by $R_{IC1}$, $R_{IC2}$.

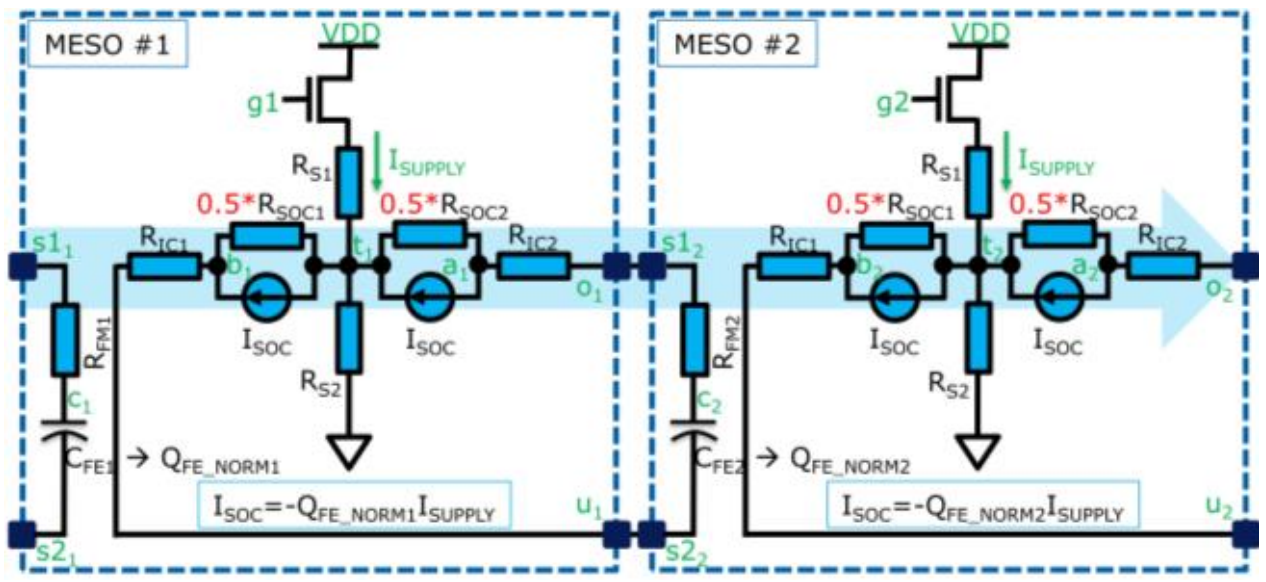

*Figure 2. The prior circuit schematic of two cascaded stages of MESO with a differential output and electrically isolated stages.*

While MESO's primary behavior is mimicked by this simplified model, the delicate dynamics and underlying physics is not properly accommodated. Particularly, the one-dimensional LK equation was used in the $C_{FE}$ element. This gives a general estimation on FE switching but did not comprehend the dynamics of the FM and AFM order parameters or their coupling. More importantly, the 3D nature of spin-charge interconversion occurs in different layers and material parameters vary across multiple layers in the whole device. Both are not modelled rigorously. This motivates this upgrade based on physics-based models.

## III. Magnetoelectric Compact Model

In the ME input part, the top ferromagnetic (FM) layer, the middle magnetoelectric (ME) layer and the bottom electrode (BE) layer essentially form a multiferroic capacitor as shown in the Fig. 3 (a). The switching of this capacitor involves both ferroelectric and magnetic dynamics, and the coupling between them. This switching stage provides the greatest contribution to the delay and thus needs to be treated accurately.

For ferroelectric dynamics, when free charge is accumulated on the FM and BE electrodes, the resulting vertical electrical field $E$ forces polarization $P$ towards pointing in the same direction. In the crystal cell, P prefers to point towards vertices of the cubic lattice. Besides, it can switch over a nontrivial two-step trajectory involving a 71° and 109 ° turning of the polarization state [18]. Therefore, polarization $P$ is not exactly parallel to the electric field $E$. Polarization in Fig. 3 (a) is shown vertical only for simplicity sake. This ferroelectric switching can be modeled with the LK equation:

$$\gamma_{FE}\frac{\partial \vec{P}}{\partial t} = -\frac{\partial(F_{bulk}+F_{elas}+F_{elec}+F_{me})}{\partial \vec{P}} \quad (1)$$

$$F_{me}(\vec{P},\vec{M}_i) = F_{DMI} = \sum_{i=1}^{n} D_{i,j}(\vec{N}\times\vec{M}_c) \quad (2)$$

where $\gamma_{FE}, \vec{P}, t, F_{bulk}, F_{elas}, F_{elec}, F_{me}$ stand for the FE damping constant, FE polarization, time, bulk FE energy, FE elastic energy, FE electric energy and magnetoelectric coupling energy, respectively [16]. Specifically, the magnetoelectric coupling energy can be calculated with Eq. (2), where $F_{DMI}, i, n, j, D_{i,j}, \vec{N}, \vec{M}_c$ are Dzyaloshinskii-Moriya interaction (DMI) energy [19], [20], index of the lattice cell of interest, number of neighboring lattice cells, index of such cell, DMI constant, Neel vector and Canted moment, respectively. Hence the $F_{me}(\vec{P},\vec{M})$ is essentially the coupling path between the ferroelectric order $\vec{P}$ and the antiferromagnetic order. In other words, this set of equations models how the applied electrical driving force will change FE order and AFM order in the ME layer.

$$\frac{d\vec{M}}{dt} = -\gamma\vec{M}\times\vec{H}_{eff} - \frac{\alpha}{M_s}\vec{M}\times\frac{d\vec{M}}{dt} \quad (3)$$

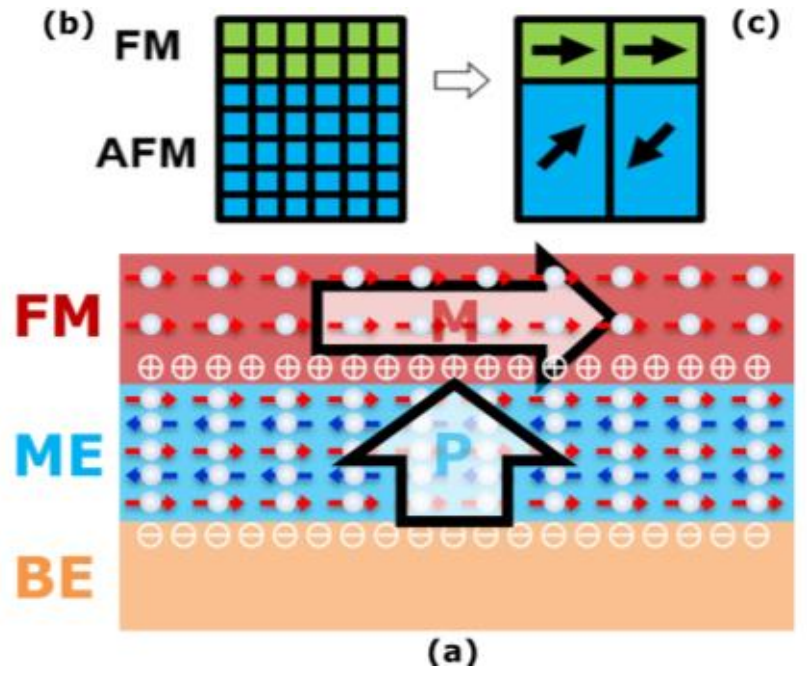

*Figure 3. The cross-section scheme of the ME input part (bottom). The evolution of the model from a discretized micromagnetic simulation domain to the macrospin compact model (top).*

$$F_{ex,int}(\vec{M}_{j,AFM},\vec{M}_{i,FM}) = J_{ex,int}\frac{\vec{M}_{i,FM}\cdot(\vec{M}_{i,FM}-\vec{M}_{j,AFM})}{\Delta_{i,j}^2} \quad (4)$$

For the magnetic dynamics of the ME part, the FM and ME layers can be meshed and simulated with a micromagnetic solver as shown in Fig. 3 (b) / (c). The governing equation for the magnetic dynamics is the LLG equation without thermal noise (3), [21], where $\vec{M}, t, \gamma, \vec{H}_{eff}, \alpha, M_s$ are the magnetization vector, time, gyromagnetic ratio, total effective field, damping constant and saturation magnetization, respectively. However, conventional micromagnetic simulation takes an excessive amount of time. To accelerate the simulation, the mesh in (b) can be approximately replaced by 4 macrospins with renormalized values of coupling between them as in (c). Due to the ferromagnetic and antiferromagnetic exchange coupling within FM and AFM layers, respectively, the top pair of spins are close to parallel, and the bottom pair of spins are close to antiparallel. The antiferromagnetic Neel vector and the canted magnetization can be obtained as following, $\vec{N} = (\vec{M}_1 - \vec{M}_2)/(|\vec{M}_1| + |\vec{M}_2|)$ and $\vec{M}_c = (\vec{M}_1 + \vec{M}_2)/(|\vec{M}_1| + |\vec{M}_2|)$. Now that the AFM order in ME layer will also interact with the magnetization in the FM layer via the interfacial exchange coupling as expressed by Eq. (4), where $F_{ex,int}, J_{ex,int}, \vec{m}_{i,FM}, \vec{m}_{j,AFM}, \Delta_{i,j}$ are the interfacial exchange energy, interfacial exchange constant, magnetization vector for FM, magnetization vector for AFM and the spin-to-spin distance, respectively [16]. With the setup above for ME part, the LK equation and LLG equation can be solved together, using material parameters as inputs.

## IV. Spin-Orbit Effect Modeled with 4x4 Matrices

In the exploration of spintronics, various advantageous device options have been proposed. Consequently, the question arises of how to simulate them in a unified framework. To address this, the charge-and-spin transport problem has been reformulated in terms of 4x4 matrices with magnetic dynamics included [6]–[8]. The elements common to various spintronics devices have been formulated and built into a set of representative modules. Hence modeling a specific spintronics device is reduced to constructing a circuit from these modules. This was termed 'modular approach to spintronics'.

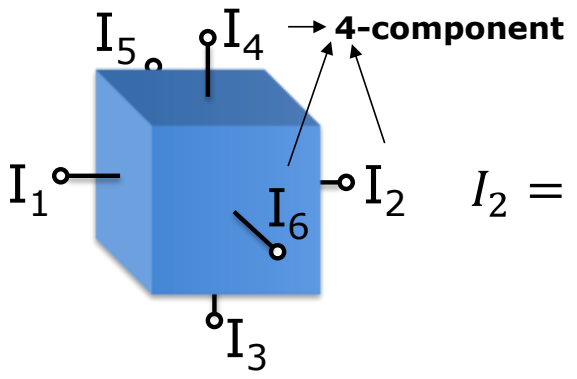

$$I_2 = \begin{bmatrix} I_2^c \\ I_2^z \\ I_2^x \\ I_2^y \end{bmatrix}, V_2 = \begin{bmatrix} V_2^c \\ V_2^z \\ V_2^x \\ V_2^y \end{bmatrix}$$

*Figure 4. The geometry of a spin-orbit coupling (SO) module in a MESO device and 4x1 vectors for current and voltage of charge and spin. Here superscript c stands for charge and superscript x/y/z stands for the electron's spin polarized in three orthogonal directions. Subscript 1, 2 … 6 are index of different facets of the cuboid.*

$$\begin{bmatrix} I^c \\ I^z \\ I^x \\ I^y \end{bmatrix} = G^E \begin{bmatrix} V^c \\ V^z \\ V^x \\ V^y \end{bmatrix} = \begin{bmatrix} G^{cc} & G^{cz} & G^{cx} & G^{cy} \\ G^{zc} & G^{zz} & 0 & 0 \\ G^{xc} & 0 & G^{xx} & 0 \\ G^{yc} & 0 & 0 & G^{yy} \end{bmatrix} \begin{bmatrix} V^c \\ V^z \\ V^x \\ V^y \end{bmatrix} \quad (5)$$

Here, the modular approach is applied to the SO output part of a MESO device. The primary physical effects are the drift-diffusion and interconversion of charge and spin currents. Fig. 4 illustrates a module for a cuboid piece of a spin-orbit-coupling (SOC) layer, which is key for the SO part of the device. To consider both charge and spin, the scalar of current and voltage at each port in regular circuit model are represented by a 4x1 vector. The superscript of c, x, y, z stand for charge and three projections of spin on the respective axes. The subscripts of 1, 2, … 6 are the indexes of cuboid facets. Hence, the overall current or voltage vector will have 24 elements, and a 24x24 tensor matrix is used to represent the conductance and dissipation of this cuboid. The generalized form of the SOC conversion is rederived from model 3 of [22]. The matrix form is shown in Eq. (5), where $I^\eta = (I_1^\eta, I_2^\eta \dots, I_6^\eta)^T$, $V^\eta = (V_1^\eta, V_2^\eta \dots, V_6^\eta)^T$ with $\eta = c, z, x, y$. Each element in the 4x4 matrix of (5) is also a 6x6 sub-matrix. Therefore, the 4x1 vector and 4x4 matrix of (5) are essentially 24x1 vector and 24x24 matrix. The elements of 24x24 matrix are functions of cuboid geometry, conductivity, spin diffusion length and spin Hall angle. Despite the ongoing discussion on the reciprocity of the forward/inverse SOC effect and the equivalence of spin Hall effect in the bulk to the Rashba-Edelstein effect at an interface [23], [24], these physics behaviors are treated with the same SOC module here. Despite their different microscopic origin, the phenomenological model considered here can be calibrated by experiment to capture these different effects.

Similar approaches have been applied to other functional modules, like ferromagnetic (FM) module, ferromagnetic-nonmagnetic interface (FM-NM) module and nonmagnetic (NM) module [8]. To account for the magnetization dynamics, terminals for magnetization orientation angles have been built for the FM module as well.

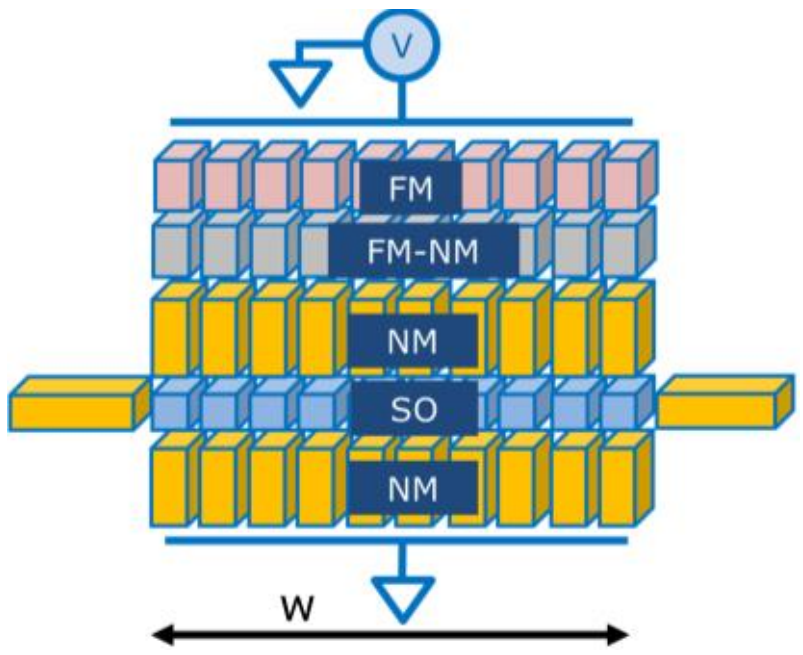

*Figure 5. Scheme for discretization of a 'slice' through the SO part into a mesh of cuboids with different categories. From top to bottom, there are layers of ferromagnetic (FM), ferromagnetic-nonmagnetic (FM-NM), nonmagnetic (NM) top, spin-orbit (SO) and NM bottom. Differential outputs use two NM modules as interconnection.*

Once these necessary modules have been prepared, we can take a vertical "slice" of the SO part, discretize, and map each cuboid to a corresponding module as shown in Fig. 5. Specifically, we have FM, FM-NM, NM, SO, and NM layers from top to bottom. Additional NM modules are placed on the left and right sides of SO to represent an interconnecting wire. By varying the mesh size one can account for the non-uniformity of quantities with varying granularity. The spin polarization in the SO part is strongly linked to the FM magnetization obtained from the ME part. During the operation of the SO slice, spin unpolarized free charge carriers (i.e. electrons with equal amount of spins up and down) could flow from the top and first traverse the FM layer. The FM forces the majority of magnetic moments (proportional to spins) of free carriers to align to the magnetization. In other words, spin as a result of polarized electrons from FM depends on the magnetization of the FM layer. Next, the partially polarized spin current encounters the FM-NM interface, where interface resistance may do additional spin filtering [25]. The charge current is conserved in the following NM layer while its spin polarization will decrease according to the spin flip length in it. Eventually, the spin-polarized current flows into the SO layer, which implements the spin-to-charge conversion. In the conductance matrix in Eq. (5), the diagonal elements stand for the intrinsic conductance for charge and for electron spins. The off-diagonal elements stand for the interaction of spin and charge. In the SO layer, a majority of the spin current is expected to get converted to horizontal charge flowing towards the differential interconnects. As a result, the charge current is generated to drive the ME part of the next MESO stage. The carriers which did not experience scattering in the SO layer continue flowing vertically through the bottom NM layer to ground.

By mapping the SO (slice) part into spintronics modules, this model can relate the material parameters to the in/out voltage/current, and the magnetization values. With this modular numerical modeling procedure, the spin-charge coupled drift-diffusion equations are solved along with magnetic dynamics.

## V. Dynamics of Combined ME and SO Parts

Using the above numerical model, initial simulation examined how a SO slice switches the ME capacitor.

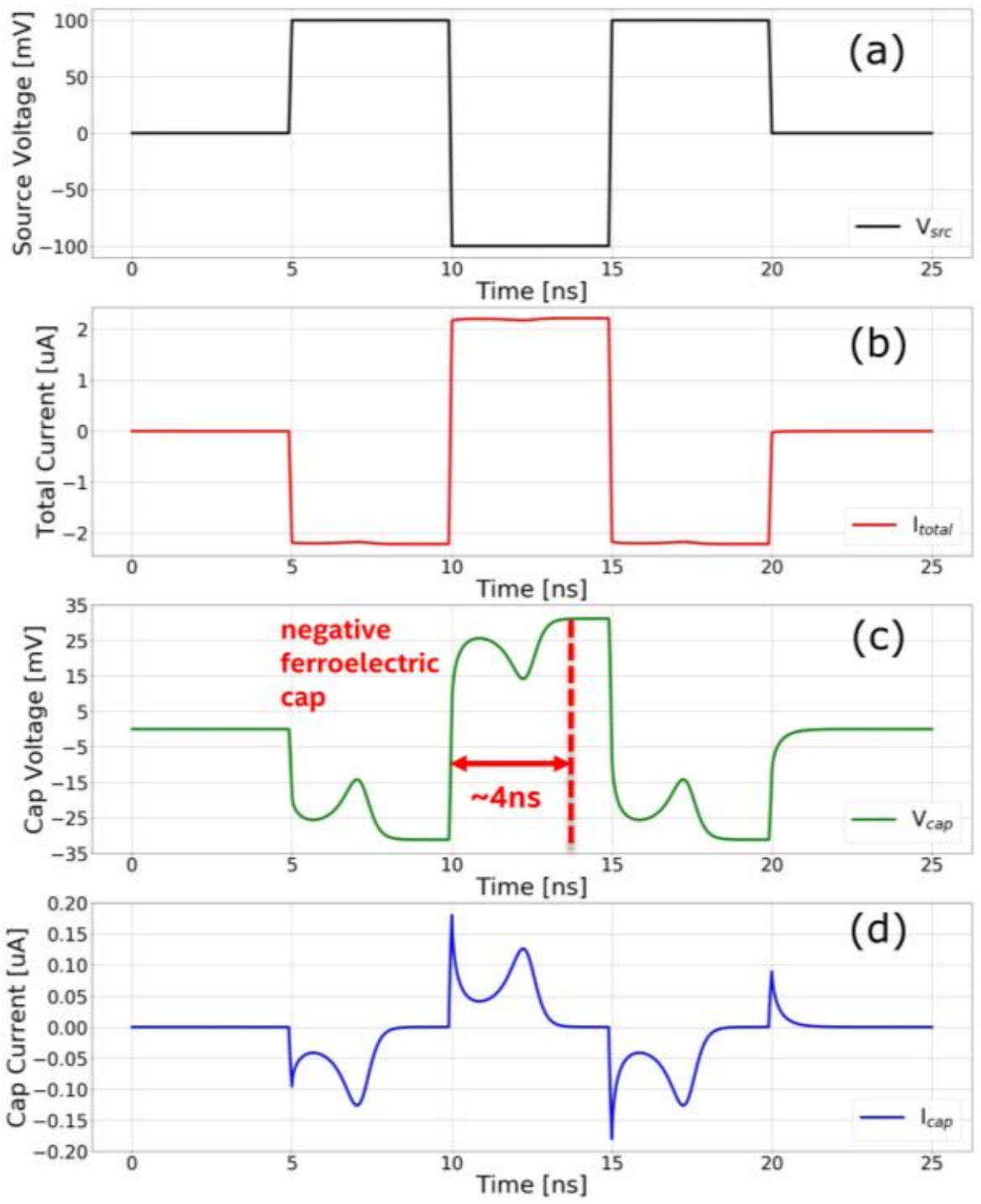

*Figure 6. Time evolution of the (a) supply driving voltage, (b) driving current in the SO part, (c) voltage at the ME capacitor, and (d) the current to the capacitor.*

To drive the SO slice, the top charge terminal is connected to an ideal voltage source Vsrc with a variable polarity. Its amplitude is 100mV, pulse width is 5ns, and rise/fall time is 0.1ns as in Fig. 6 (a). In graph (b), the vertical charge current, Itotal, flows between Vsrc and gnd, and its amplitude is ~2.2uA. Graph (c) shows the charge voltage Vcap across the ME capacitor electrodes. Each time Vsrc changes polarity, the resulting Vcap will switch its polarity as well. During each ramp up, the absolute value of Vcap would increase, decrease, and increase again to saturation. This is caused by the transient negative capacitance [26]. The saturation level of Vcap is ~31mV, which is above the assumed coercive voltage ~20mV in ME material. This Vcap saturation level is essentially determined by the product of Isoc and Rsoc, which stand for inverse SOC current and SO lateral equivalent resistance. It takes about 4ns for Vcap to saturate. The varying charge current through the ME capacitor, Icap, is shown in Fig. 6d. The nonlinear behavior also stems from the transient characteristics of ferroelectric switching. Once the Vcap saturates, the Icap reduces to zero, which means the charging of capacitor is complete. The peak amplitude of Icap is ~0.2uA, provided that the inverse SOC current conversion efficiency is ~10%.

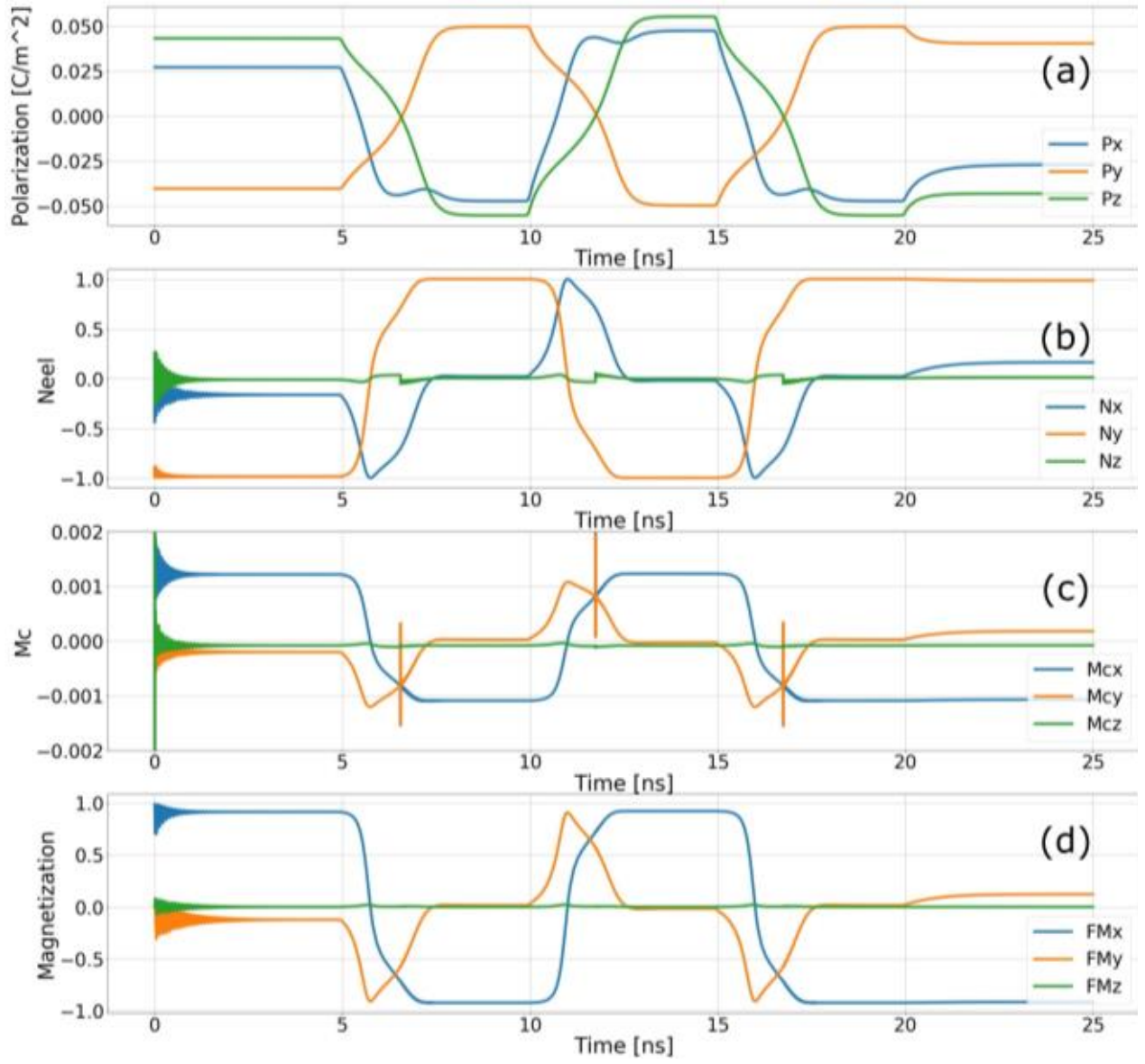

*Figure 7. Time evolution of the (a) the ferroelectric polarization, (b) the antiferromagnetic Neel vector, (c) the canted magnetization and (d) the magnetization in FM electrode in ME layer. Each variable is a vector and projected onto x/y/z directions.*

In Fig. 7, the time evolution of FM magnetization, FE polarization in ME, AFM Neel vector in ME and canted magnetization $M_c$ in ME, are shown. For graph (a), the polarization switches between two opposite orientations, +x-y+z and -x+y-z. This is because the polarization is aligned from the center to cube vertices in the crystal lattice cell. Each time when Vsrc and Vcap switch polarities, the FE polarization will change its sign if Vcap exceeds the FE coercivity. In graph (b), the AFM order will follow the pace of polarization, which makes the Neel vector to alternate between positive and negative y-directions, which is the in-plane easy axis. In graph (c), the canted magnetization $M_c$ switches between positive and negative x directions, which is the in-plane easy axis. In graph (d), with the coupling to $M_c$, the FM magnetization aligns in the same direction. Hence the main projection of FM is on the x-axis.

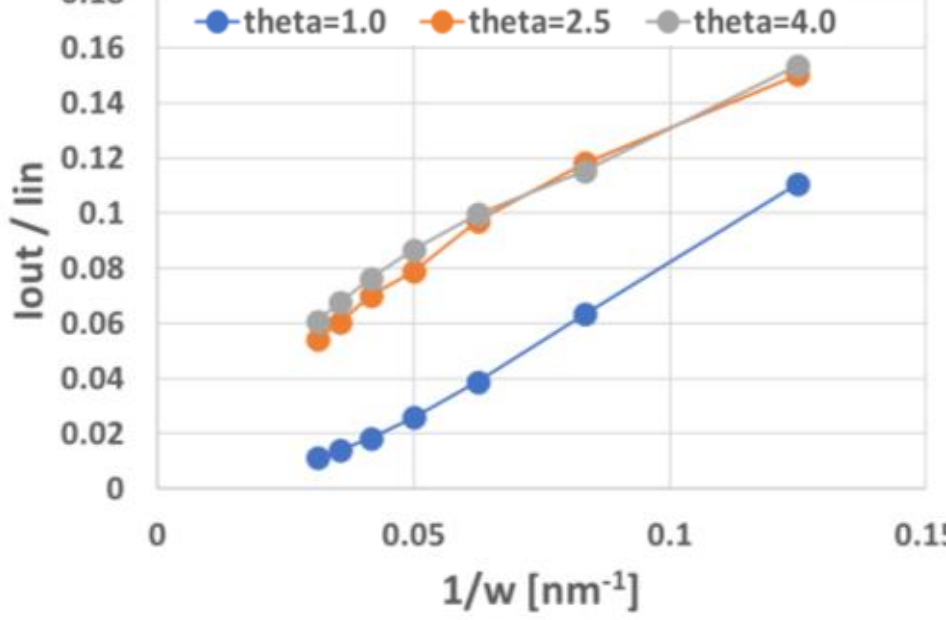

*Figure 8. Dependence of the ratio of the output current to the input current on the SO slice width at various values of the spin Hall coefficient*

With this combined ME-SO simulator, various parameter dependencies can be explored. In the following, we present one example where the width of the SO slice is varied to understand its scaling behavior. In the setup of Figs. 6 and 7, the width of the SO slice per one output wire, *w*, is 20nm (2nm x 10 columns as in Fig. 5). By tuning the number of cube columns in the SO slice and repeating the same simulation, a series of results was generated. The ratio of *Icap* and *Itotal* (also the *Iout* and *Iin*) of the SO slice has been obtained and plotted vs. 1/*w*. Using the spin Hall angle, *Θ*, of 1.0, 2.5, and 4.0, similar curves were obtained and plotted in Fig. 8. The simulation results show that current conversion efficiency of the SO slice, Iout/Iin, has a quasi-linear dependence on 1/*w*. In other words, with narrower *w* of the SO slice, a higher Iout/Iin leads to more output current and makes the ME capacitor able to be charged faster with the same current from power supply. This would continue to be beneficial with further scaling down of the SO width. Comparison of different *Θ* values also confirms that the current conversion efficiency is higher for larger *Θ*.

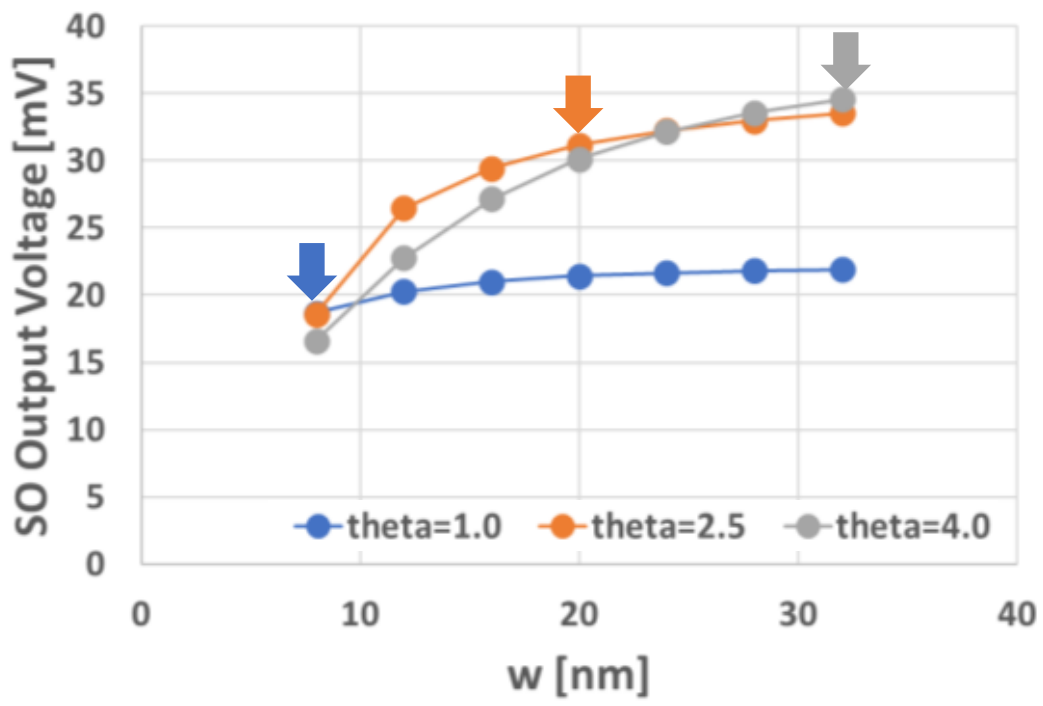

*Figure 9. Dependence of the output voltage from the SO part on the magnet width at various values of the spin Hall coefficient.*

Similarly, the maximum of Vcap absolute value is obtained as a function of *w* as shown in Fig. 9. With smaller *w*, the *Vcap* decreases too, which is mainly caused by the decrease of the internal resistance of the current source $R_{ISOC}$. This behavior is more obvious for the case with larger spin Hall angle. To ensure the normal functionality of the MESO device/circuit, the absolute value of *Vcap* needs to be larger than the coercive voltage of the ME material. To address this requirement, when *w* of SO further scales down, either the decrease in the output voltage needs to be compensated by another factor, or the ME input coercive voltage needs to be decreased by the material optimization. Comparing curves at different *Θ* values, we notice that the saturation level of *Vcap* increases with larger *Θ*, but this increase also gradually saturates. At different *w* values (labeled by down arrows), the optimal *Θ* for higher *Vcap* is different. This suggests that the spin Hall effect optimization should go hand in hand with the device scaling: at relatively large *w*, higher *Θ* is preferred; while for smaller *w*, *Θ* requirement would be relaxed.

## VI. 3D Effect in the Spin-Orbit Part

So far, an individual MESO device was modeled with the complete ME part and a single SO slice. Further study suggests that there are nonzero charge/spin currents in the direction perpendicular to the output wires. Hence, a full model, where SO part comprises multiple SO slices in parallel, becomes necessary to handle the 3D nature of SO effects as in Fig. 10.

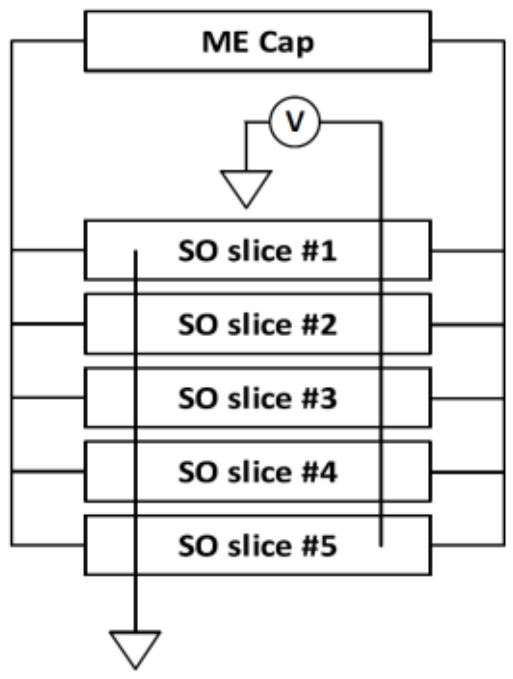

*Figure 10. Circuit schematic of the 3D structure in the SO part via parallel vertical "slices". The top/bottom electrodes of all SO slices are merged and connected to the power supply and to ground, respectively.*

Fig. 11 shows the simulation results for five SO slices used to switch the ME part. The corresponding terminals of five SO slices are merged to form a full SO output. The same ideal voltage source as in section V is used for the driving force. In graph (a), the voltage waveform has the same amplitude of 100mV and 5ns pulse width. In graph (b), the amplitude of current from power supply increases from 2.2uA to ~9uA. This is mainly because more SO slices in parallel reduce the resistance between the power supply and the ground. In graph (c), the *Vcap* across the ME capacitor shows similar behavior as before. But there are two interesting changes. One is that the saturation level of *Vcap* increases from 31mV before, in section V, to 40mV now. This indicates that the multi-slice interaction leads to increase of SOC output voltage. Intuitively, when the spin current is injected into SO layer, the inverse SO effect creates charge currents towards all surfaces of the SO layer: not just left and right, but also the front and back surfaces. This results in the potential differences between these surfaces. The backflow currents caused by these potentials are experiencing the direct spin Hall effect and add to the injected spin current. Modeling more SO slices allows us to capture these 3D effects.

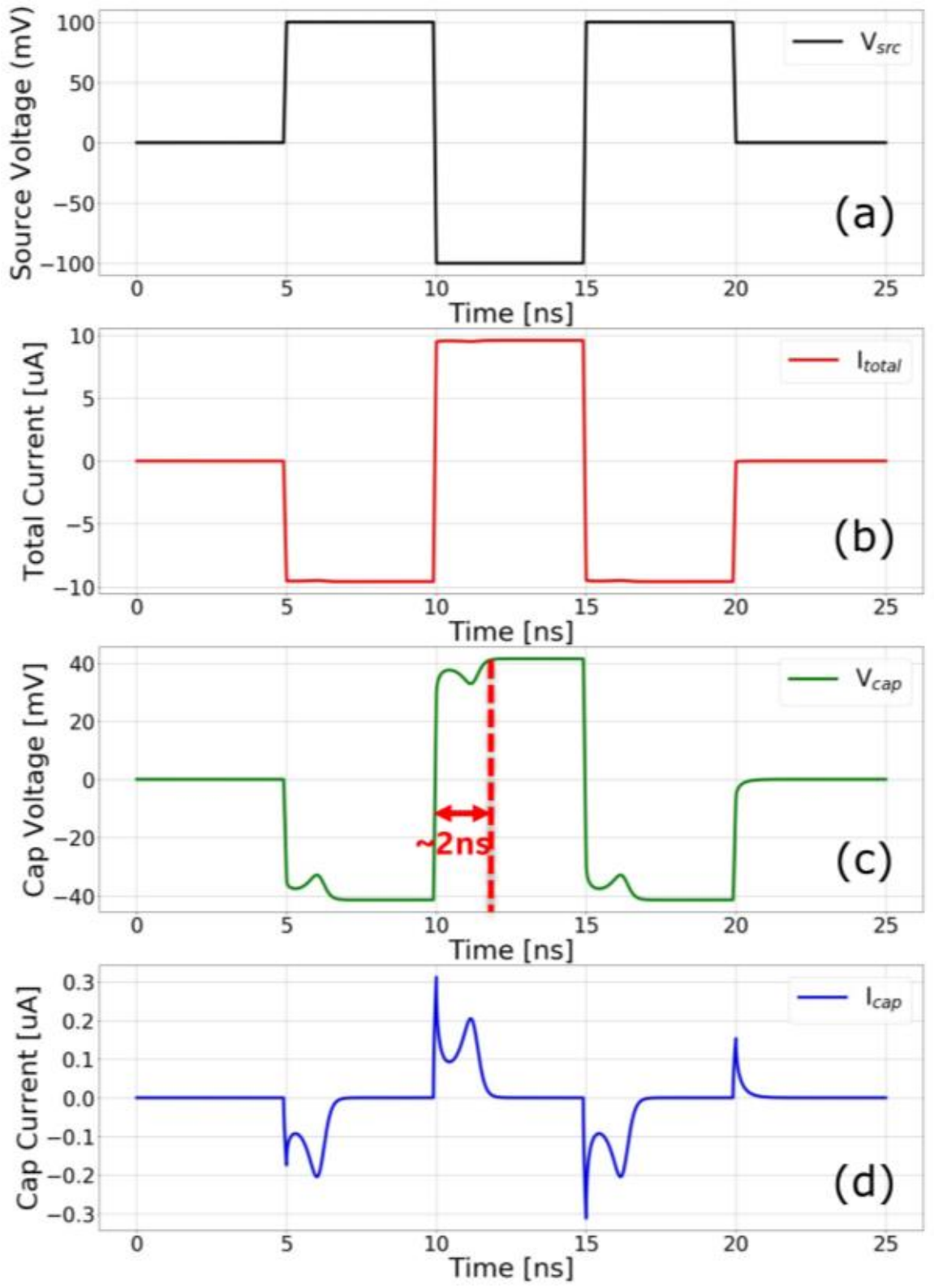

*Figure 11. Five SO slices driving the ME capacitor. Time evolution of (a) power supply VDD (b) power supply current (c) voltage across ME capacitor (d) charging current for ME.*

The other change is the time it takes for *Vcap* to saturate, which is mainly attributed to differences in *Icap*. In graph (d), we see that the charging current, *Icap*, also increases with more SO slices included. As a result, charging the ME capacitor takes a shorter time ~2ns instead of ~4ns with a single SO slice.

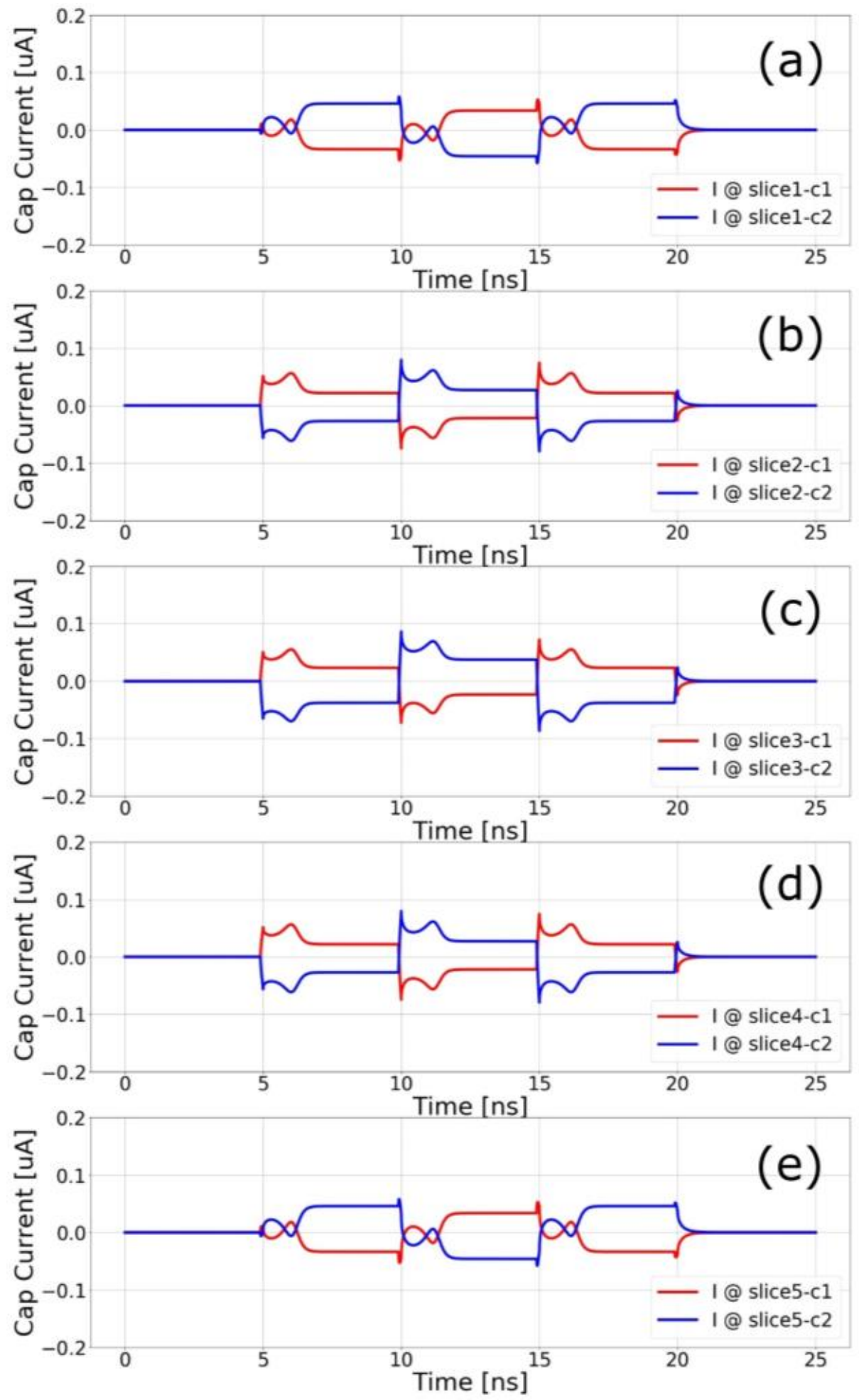

*Figure 12. Time evolution of lateral charge current flowing into / out of the differential output terminals for each SO slice.*

Fig. 12 shows the lateral currents flowing into / out of the differential output of each SO slice in the 3D MESO model. Each graph corresponds to one SO slice, and the two curves are measured on each of the output charge terminals. Hence, in each graph, the two curves have the same amplitude and opposite signs. It can also be noticed that the currents in these 5 SO slices are not identical. Slices show symmetric behaviors with respect to the central slice. For instance, the slice (a) is similar to (e), and slice (b) is similar to (d). To validate that this is caused by the inter-slice interaction, the slice-to-slice interconnection was cut off in a control simulation (not shown here); that made each slice reduce to being in the same state.

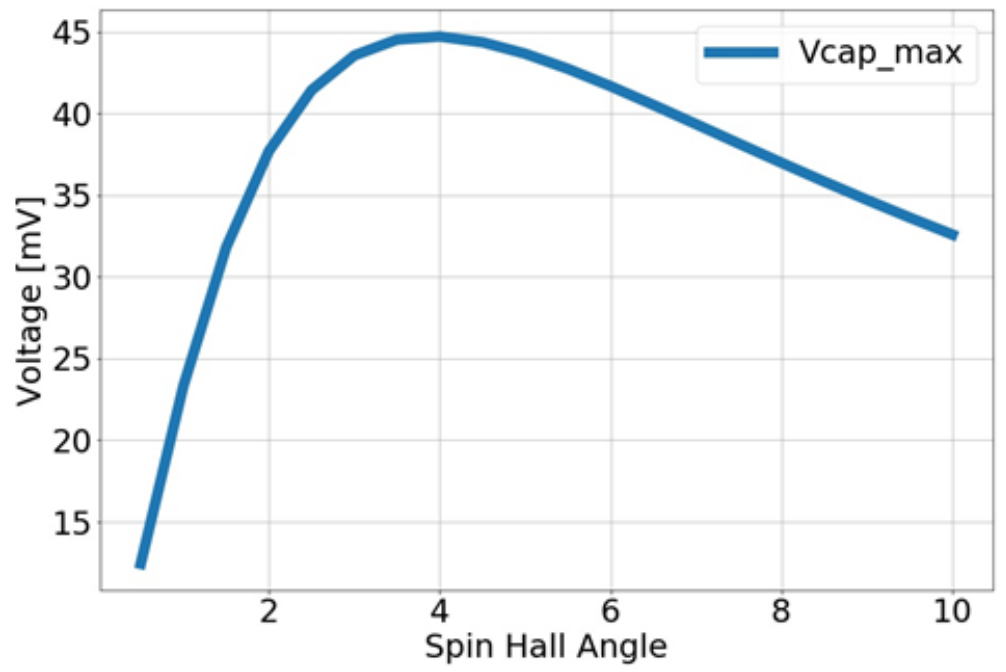

*Figure 13. SO part differential output voltage as a function of the spin Hall angle (Θ) in SO layer.*

With the full 3D MESO model, sensitivity study of several parameters is enabled. Here we revisit the dependence on a major factor, spin Hall angle ($\Theta$). As shown in Fig. 13, the $\Theta$ is swept from 0.5 to 10 [27]. For each $\Theta$ value, similar switching simulation was done to extract the maximal SO output voltage or *Vcap*. Previously, this *Vcap* was expected to increase with larger $\Theta$. In contrast, the simulation results show a surprising non-monotonic trend, where the *Vcap* vs. $\Theta$ first increases, then saturates, and finally decreases. The optimal $\Theta$ value is ~4 and as shown in the earlier discussion, it varies with both geometric and material parameters. In the simplified SO 'model 1' of [22], the open circuit charge voltage for the inverse spin Hall effect can be described by Eq. (6). The $V_{so}, \theta, \sigma, t, \lambda, I_3^z$ are charge output voltage, spin Hall angle, conductivity, thickness, diffusion length and injected spin current into the SO layer. In Eq. (6), $\Theta$ is present in the numerator and its 2nd power is in the denominator as well. The general explanation for this behavior is that once spin current is injected vertically, the left/right surface would have a charge voltage built up. This accumulated charge voltage would inversely create an opposite vertical spin current to resist the injected spin current. As a result, the net injected spin current would be lower, and the output voltage would reduce as well.

$$V_{so} = (V_1^c - V_2^c)|_{I_1^c = I_2^c = 0} = \frac{\theta}{2\sigma\theta^2 + \sigma\frac{t}{\lambda}\coth\frac{t}{2\lambda}} \frac{I_3^z}{w} \tag{6}$$

## VII. MESO Symbol

We proceed analyzing MESO circuits using the 3D physics based compact models of MESO as elements in larger circuits. For this the models of the ME part and the SO part used in previous sections are wrapped into the circuit symbols as shown in Fig. 14. These ME and SO symbols are connected via the pins 'theta' and 'phi' to transmit the values of the magnetization orientation angles. For the ME part, the differential input nodes, n1 and n2, are the top and bottom electrodes of the ME capacitor. The SO symbol provides the charge and spin c/z/x/y pins from left, right, top, and right surfaces. The spin related terminals are grounded, and the charge terminals can connect to other charge-based components.

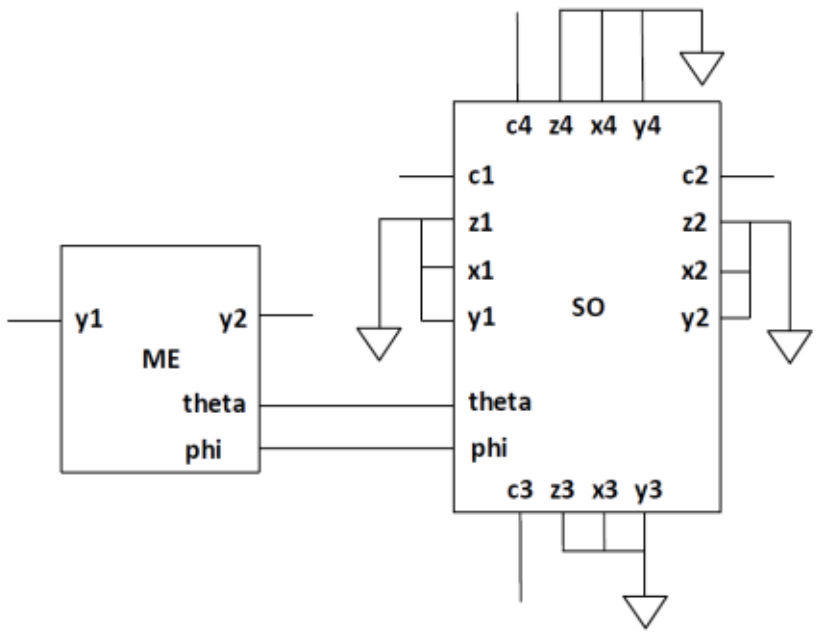

*Figure 14. Circuit schematic of ME part and SO part within the MESO 3D device symbol.*

## VIII. MESO Majority Gate

A primary circuit that MESO can implement efficiently is the majority gate. Instead of using more than 10 CMOS transistors to implement a 3-input majority gate, MESO logic only needs one MESO device with one NMOS transistor. Besides, with a single MESO device as inverter logic, majority gate would form a complete logic family and enable the design of any arbitrary circuit logic function.

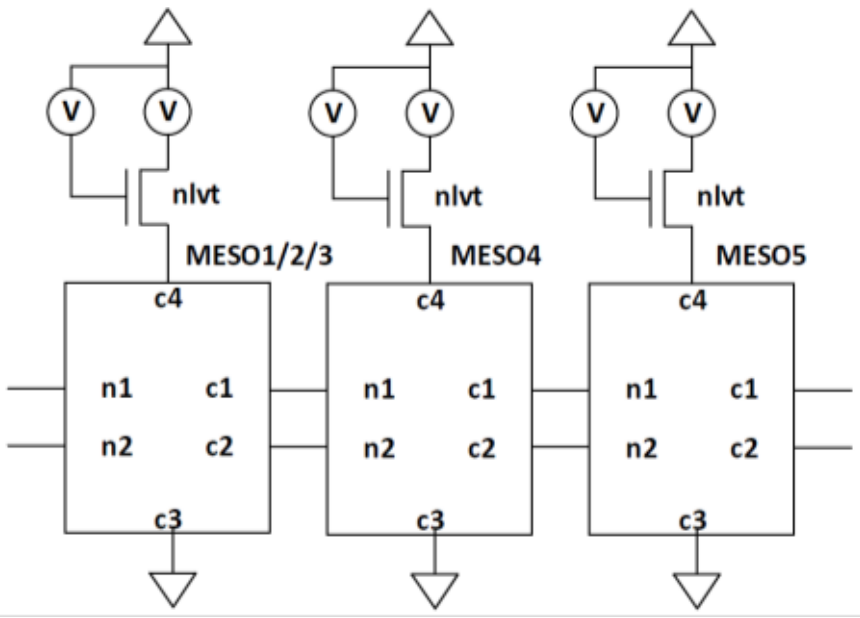

*Figure 15. Circuit schematic of a 3-input MESO majority gate.*

In Fig. 15, MESO symbols are used to construct such a 3-input majority gate. It should be noted the three-input MESO #1, 2, and 3 have their c1 and c2 terminals merged, respectively. The merged c1 and c2 then connect to the middle MESO #4, which serves as a minority gate function. Using MESO #5 as the load and an inverter as well, the minority gate is converted back into the majority gate. The n1/n2 terminals of the 3-input MESO are floating as the ME capacitor maintains its state. The c1/c2 of MESO #5 are floating as the ME input is isolated from the SO output. To control this majority gate, two non-overlapping clock signals are needed, where the first clock controls MESO #1, 2, and 3 and the second clock controls #4. For MESO #5, the access transistor and voltage sources are plotted for consistency but are not necessary for the operation.

In Fig. 16, a case with specific initial conditions is shown as an example. In graph (a), the VG1 and VG2 are two non-overlapping clock signals applied to MESO #1, 2, 3 and #4, respectively. The amplitude is 0.85V and the pulse width is 5ns. Delay from VG1 to VG2 is 6ns and rise/fall time is 0.1ns. When VG1 becomes high, the MESO #1, 2, and 3 will be enabled and generate a nonzero voltage Vcap4 at the terminals n1/n2 of MESO #4. The voltage is shown in graph (b), with amplitude of 31mV and pulse width of 5ns. This voltage will update the ME capacitor state in MESO #4. When VG2 becomes high, only MESO #4 is enabled, which generates a nonzero Vcap5 across n1/n2 of MESO #5 as shown in graph (c). This voltage will update the ME capacitor state in MESO #5 and complete the majority gate function. When the first pulses of VG1/VG2 are on, the Vcap4 and Vcap5 show transient characteristics of negative capacitance, which is the signature of the ME capacitor switching. For the following VG1/VG2 pulses, despite the nonzero Vcap4 and Vcap5, the switching is not shown and the already switched ME capacitors are just saturated and relaxed again.

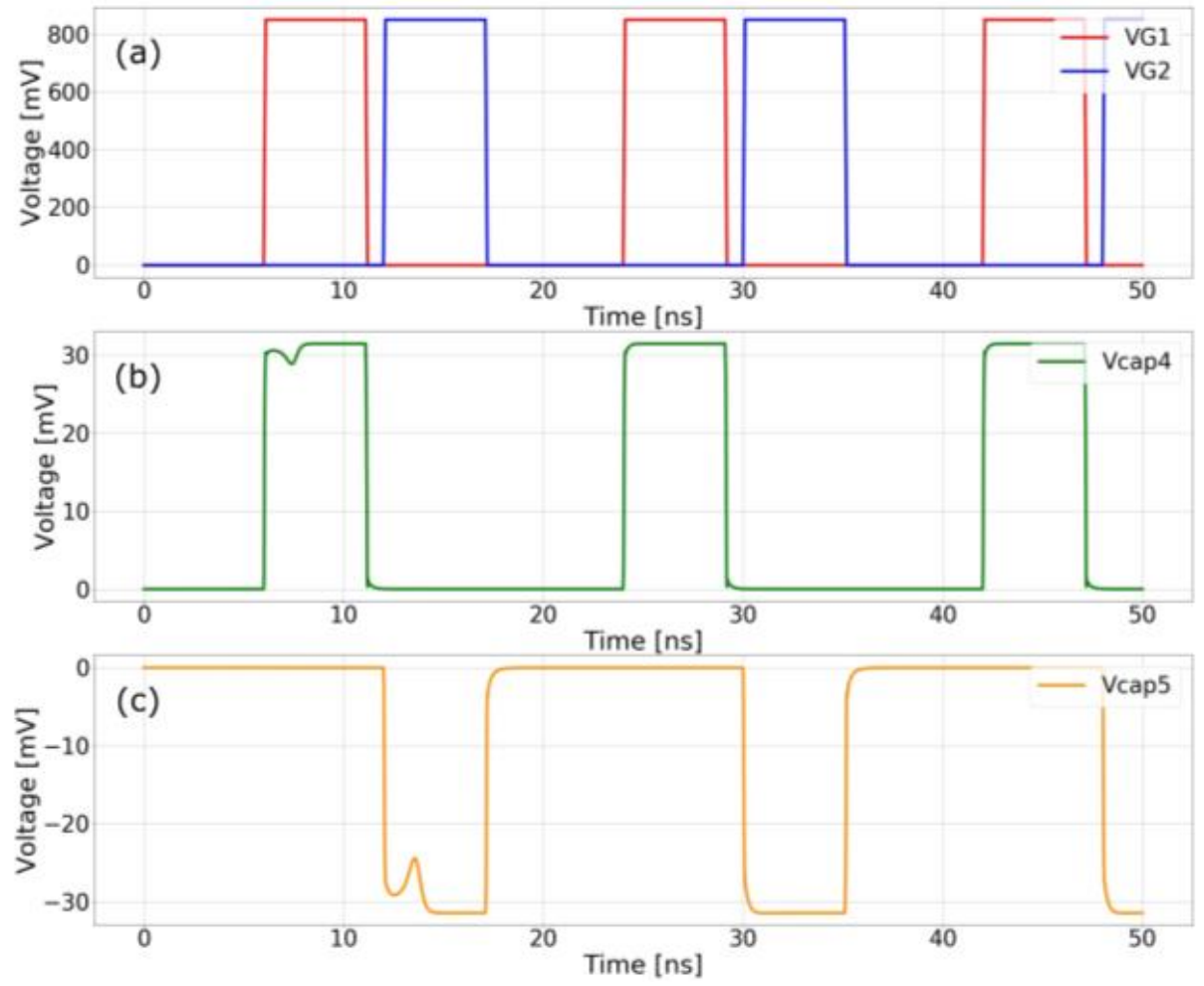

*Figure 16. Time evolution of clock signals and Vcap voltage for 3-input MESO majority gate.*

In Fig. 17, the corresponding magnetization projection, FMx, along the in-plane easy-axis direction is plotted versus simulation time for each of the MESO stages. In this case, the MESO #1, 2, 3, 4, and 5 have been initialized as -1, -1, -1, -1, and +1, respectively. -1 or +1 refers to the sign of the FMx. Throughout the procedure, the MESO #1/2/3 state remain the same as graph (a). Since three input MESO #1, 2, and 3 are all -1, their minority state is +1. Hence MESO #4 is driven to switch from initial -1 state to this minority state +1 at around 6ns. This realizes the minority-gate function. As shown in the MESO ring oscillator section (in the supplementary material), each MESO stage can serve as an inverter. Since MESO #4 switches to +1 state, MESO #5 would further invert from +1 to -1 state. The -1 state in MESO #5 is essentially the majority state of the MESO #1, 2, and 3.

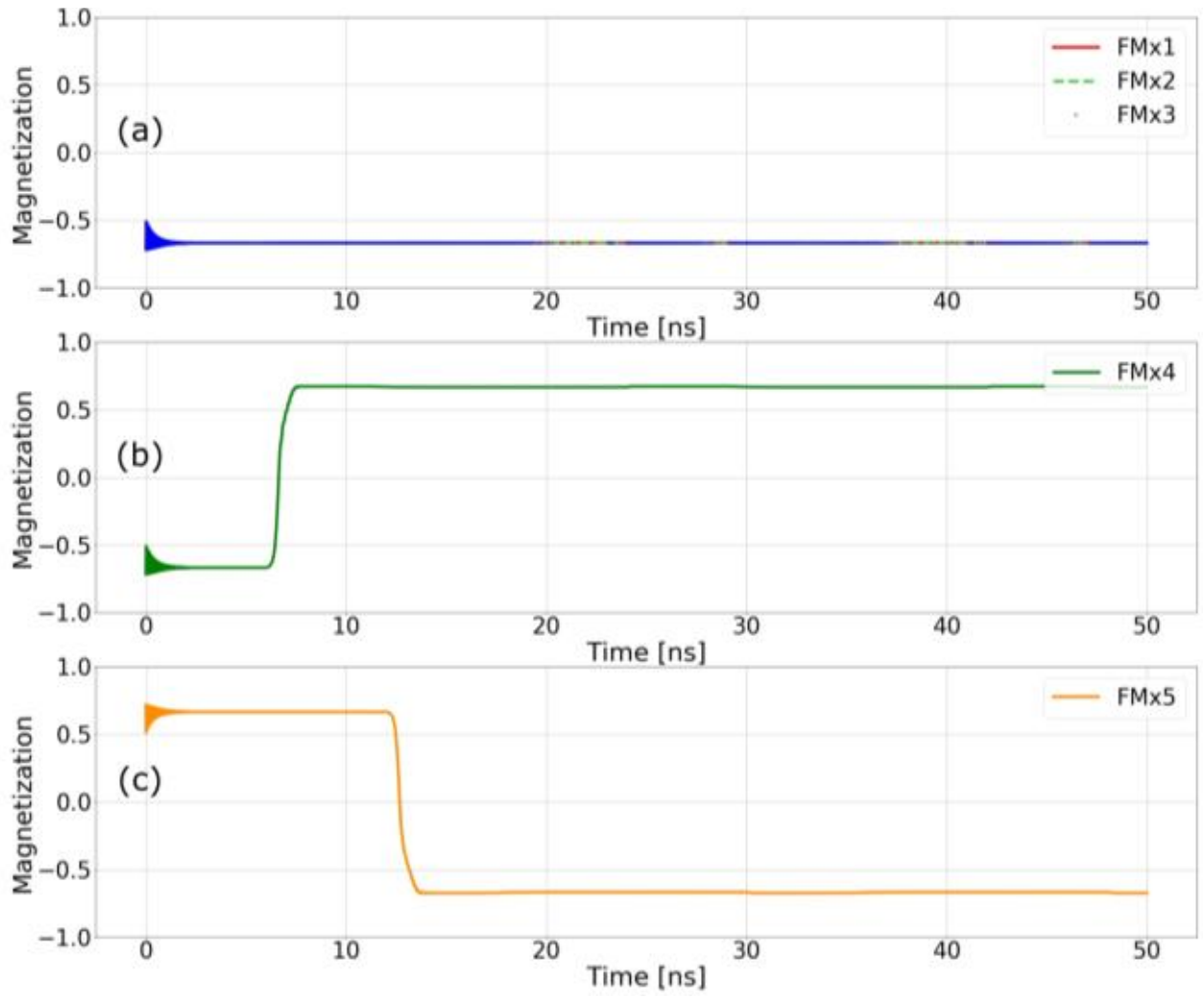

*Figure 17. Time evolution of FM layer magnetization projected along in-plane easy-axis direction in each MESO device of 3-input majority gate.*

This provides the initial demonstration for 3-input majority gate with the physics-based MESO model. To exhaustively validate the 3-input MESO majority gate, the cases with different initial conditions have been simulated and results were postprocessed with python script. For 5 MESO devices, there are $2^5$=32 possible initial conditions. The magnetization projection of MESO devices #1 to #5 and the Vcap for MESO #4 and #5 versus time, were simulated. All the magnetization for the input MESO devices #1 to #3 remain unchanged throughout the simulation. For MESO #4, some cases switch, but several cases turn out to have incomplete switching. By checking the Vcap of MESO #4, it is found that the Vcap absolute value for these half-switched cases are ~10mV. Since the ME capacitor has a coercive voltage of 20mV, the switching cannot occur. Meanwhile, one aspect in common for these unswitched cases is that the input MESO stages have “ABB” patterns, like +1/-1/-1, or +1/-1/+1 and so on. We further find that the current between the converged joint nodes c1/c2 have made the SO output voltage lower than that of an individual MESO driving device. Like the MESO behavior in the ref. [15] circuit study, with a larger number of inputs, there is expected to be more reduction of voltage at the differential output.

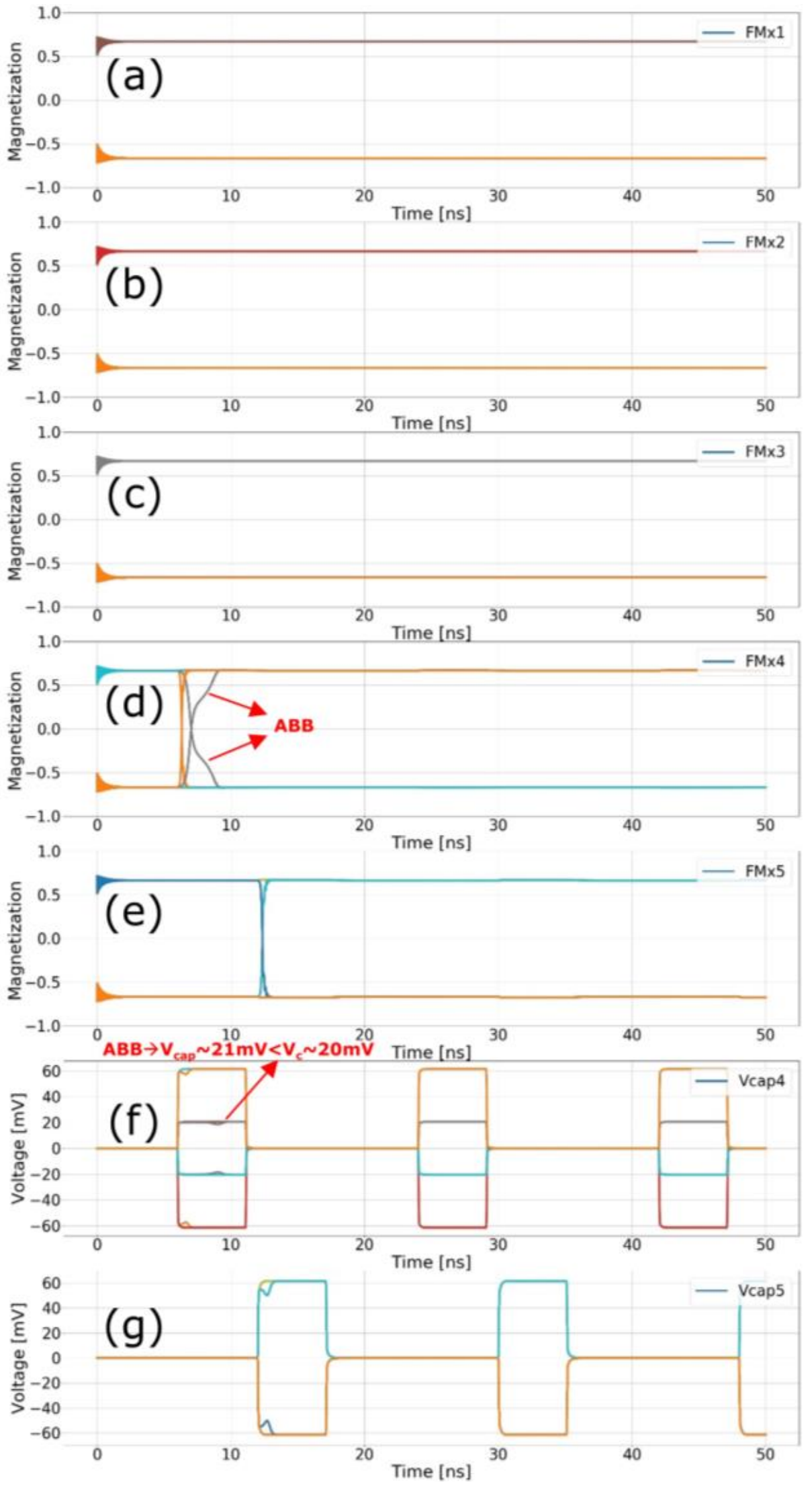

*Figure 18. Time evolution of magnetization along in-plane easy-axis direction (a-e) and Vcap of MESO #4/5 in a 3-input majority gate for all $2^5$=32 initial conditions. The power supply voltage is increased to 200mV and all the cases are functional.*

To solve this issue of smaller input signal, the larger the fan-in of a majority gate, and to ensure this majority gate can operate with various initial conditions, the power supply has been increased from 100mV to 200mV. As shown in the Fig. 18, with higher VDD, all the cases with different initial conditions can have MESO #4 and MESO #5 switched completely like graph (d-e). For MESO #4, some cases still show longer switching time, which are also caused by the “ABB” input pattern and shrinkage of Vcap. From the Vcap of MESO #4 and #5 in Fig. 18 (f-g), it can be noticed that the saturation level of voltage still has two groups, which are around 50+ and 20+ mV, respectively. The 20+mV Vcap cases are also the ones with “ABB” input patterns. However, by increasing the VDD level, all the logic functionalities for this 3-input majority gate are realized and validated.

In larger scale design and realistic tape out, the reduction of Vcap with more input for the majority gate is indeed a critical factor to consider. With the device and process variations, this needs more margin for tolerances. While the issue here is roughly solved by increase of the supply voltage, it costs more energy consumption. More importantly, it indicates that more optimization of materials for both the ME part and the SO part is necessary. For the SO part, better spin-charge conversion efficiency needs to be explored to enhance both output voltage and current levels. For the ME part, lower coercive voltage and coherent switching is critical for normal logic function and lower energy switching.

## IX. Insights into MESO Operation from Simulations

In this work, a multi-physics-based model is built for the MESO logic device. The FE/AFM/FM material systems are seamlessly integrated and simulated in the SPICE circuit simulation environment. This closes the gap between the material, device, and circuit simulation. With this methodology, the design space could be explored to better understand the MESO behavior, such as geometric scaling, 3D non-uniform charge-spin conversion as well as the non-monotonic dependence on spin Hall angle. The ring oscillator and the majority gate circuits are designed and validated by simulation. The strong correlation between the device metric and circuit behavior can be observed, which will guide future optimization.

## Acknowledgment

The authors gratefully acknowledge fruitful discussions with Manuel Bibes, Felix Casanova, Ramamoorthy Ramesh, Lane Martin.

# Supplementary Material for the paper "Physics-Based Models for Magneto-Electric Spin-Orbit Logic Circuits"

## X. MESO Ring Oscillator

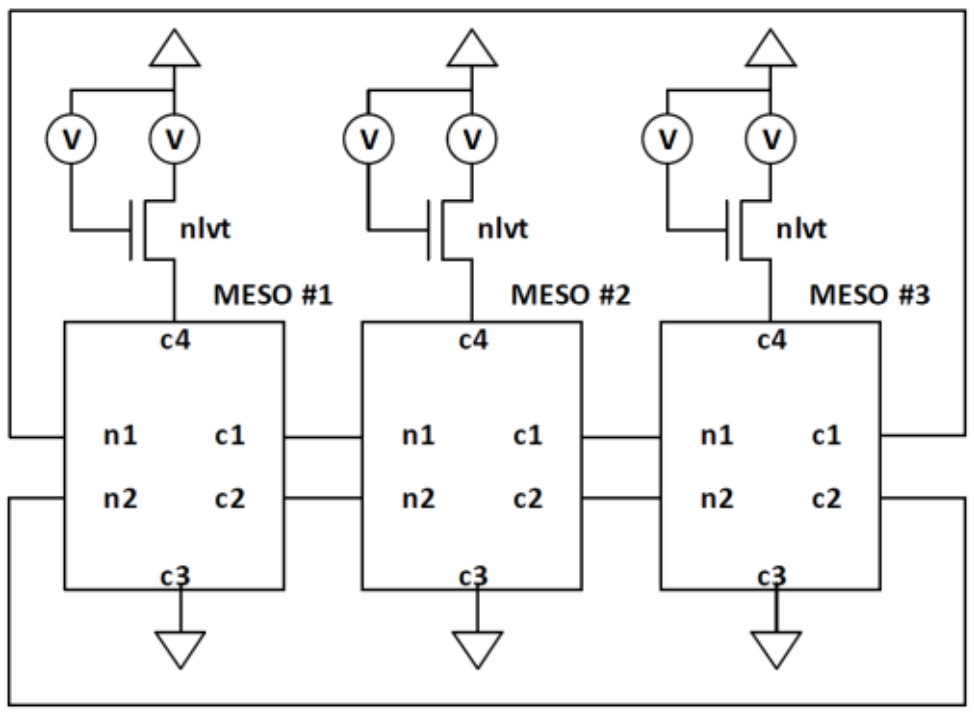

*Figure 1. Circuit schematic of a MESO ring oscillator.*

Using the MESO 3D device symbol, an essential test structure for logic, the ring oscillator, can be constructed as in Fig. 1. With differential input and output of each MESO, the c1/c2 terminals of the previous stage can connect to the n1/n2 terminals on the next stage, respectively. Hence, three MESO devices are sequentially cascaded in this manner. The output of MESO #3 connects back to the input of MESO #1 to form a loop. Each MESO stage is clocked by a low-threshold-voltage NMOS transistor from the Intel 22nm low-power FinFET technology. The drain node is held at 100mV constantly and the gate of each transistor is controlled by an individual clock with the amplitude of 0.85V. When one clock signal goes high, the correspondent MESO will be enabled to drive the next MESO.

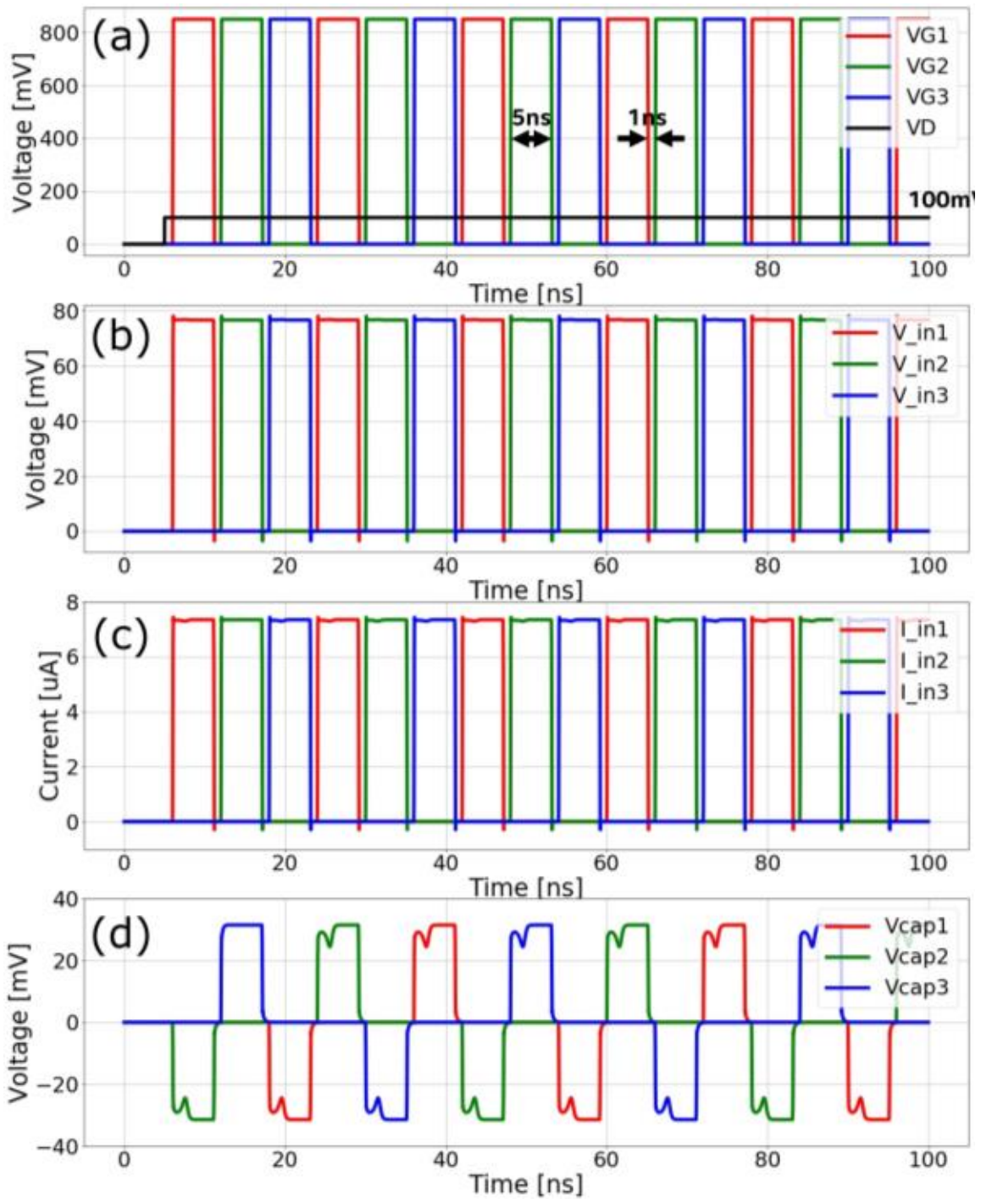

*Figure 2. Time evolution of (a) clock signals VG, power supply VD (b) input voltage at c4 (c) input current at c4 (d) SO output voltage Vcap in a 3-stage MESO ring oscillator.*

Fig. 2 shows the simulation results from the 3-stage MESO ring oscillator. In graph (a), the voltage applied to the drain and gate terminals of each transistor. The drain node voltage, VD, becomes high after 5ns and has the rise time of 0.1ns. The clock signals, VG1/VG2/VG3, are applied to the gate of the corresponding transistor. The clock period is 18ns with the pulse width of 5ns and the delay between adjacent clock pulses is 6ns. In graph (b), the resulting input voltages at c4 node of each MESO are plotted. The pulse width and period of these input voltages follow the same pattern as the clock signals. Because of the voltage drop across the transistor channel, the amplitude of input voltage is ~78mV. In graph (c), the vertical current from transistor to c4 node is also shown, which has amplitude ~7.5uA. In graph (d), generated SO part output voltages, *Vcap*, are plotted. When VG1 becomes high, the MESO #1 is enabled, Vcap2 becomes nonzero and MESO #2 switches. When VG2 is enabled, MESO #2 generates a nonzero Vcap3 to switch MESO #3. The similar procedure repeats and the *Vcap* of each stage keep alternating their signs by turns, which eventually creates the oscillation behavior. It can be noted that the amplitude of *Vcap* is ~31mV instead of 40mV previously, since input voltage/current at c4 node is lower now. Every time the Vcap curve switches polarity, consistent characteristics of transient

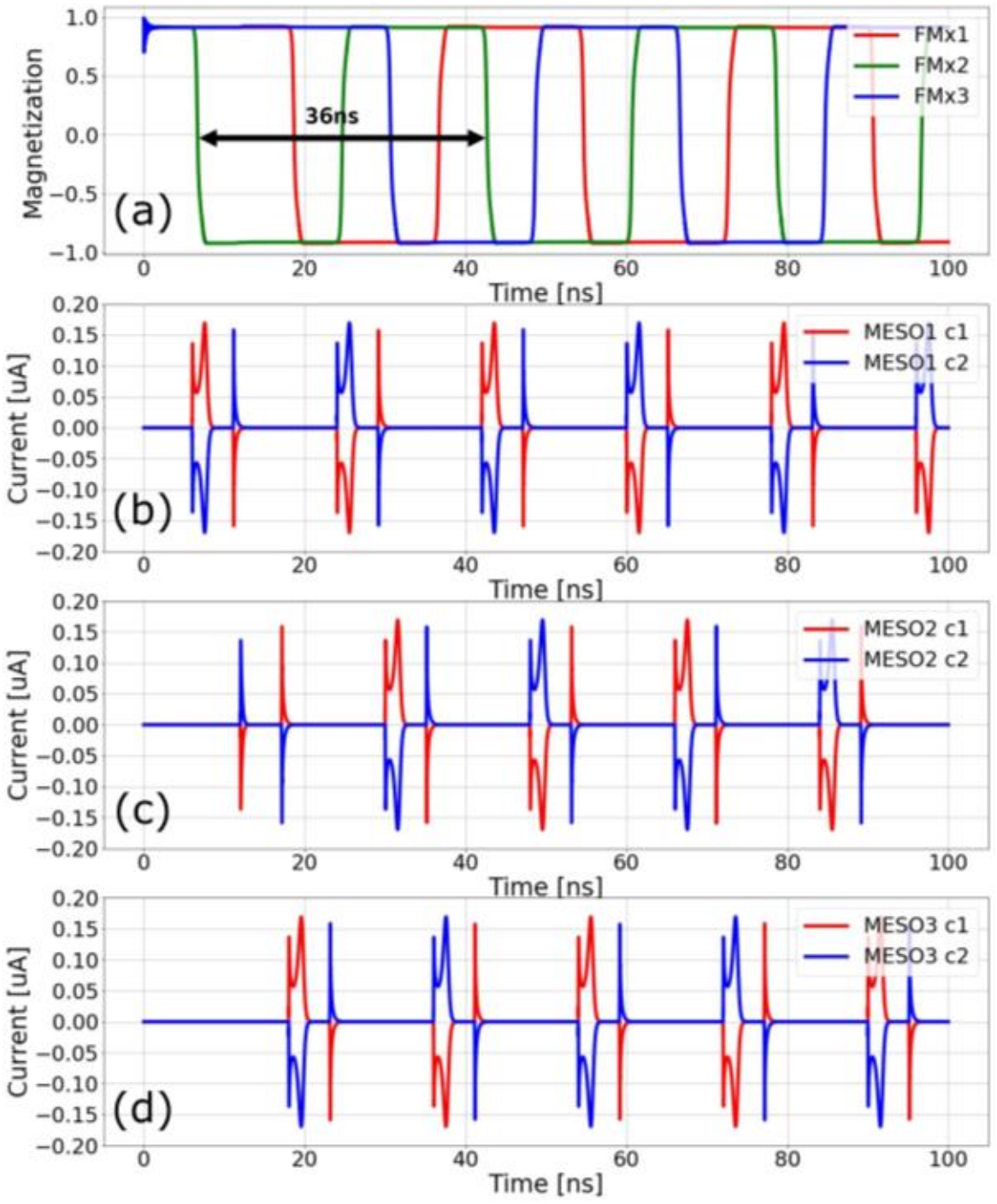

*Figure 3. Time evolution of magnetization projection and charging current in a MESO ring oscillator.*

negative capacitance can be seen.

In Fig. 3, the time evolution of FM magnetization and charging current for each MESO stage are shown. One critical advantage of MESO is the nonvolatility, where the information state is kept by the state of FE/AFM/FM coupled system. Conversely, the magnetization state can be extracted to represent the information state of MESO. In graph (a), the FM magnetization projection along in-plane easy-axis is plotted. Since MESO stages switch in turns, the magnetization also flips the directions one-by-one and the period was found to be ~36ns from edge to edge. In graph (b-d), the charging current on c1/c2 of each MESO is collected. It can be observed that each time MESO #N is enabled, the current from c1/c2 will become nonzero to charge the ME capacitor in MESO #N+1. Specifically, at the rising edge of VG, the current shows two-peak profile for charging. At the falling edge of VG, the current shows one-peak profile since the polarization of ME will restore from saturation to remanent status. It should be mentioned that in this case the period can be further compressed and only limited by the switching speed of a single stage.